\documentclass[useAMS,usenatbib]{mn2e}
\usepackage{graphicx}
\usepackage{natbib}
\usepackage[T1]{fontenc}
\usepackage{aecompl}
\usepackage{color}
\usepackage{amsmath}
\newcommand{\comments}[1]{}

\newcommand\aj{{AJ}}   % Astronomical Journal
\newcommand\apj{{ApJ}}   % Astrophysical Journal
\newcommand\apjl{{ApJ}}   % Astrophysical Journal, Letters
\newcommand\aap{{A\&A}}   % Astronomy and Astrophysics
\newcommand\mnras{{MNRAS}}   % Monthly Notices of the RAS
\newcommand\pasp{{PASP}}   % Publications of the ASP
\newcommand\memsai{{Mem.~Soc.~Astron.~Italiana}}  % Mem.Societa Astronomica Italiana
\title[HST+COS spectra of the double white dwarf CSS~41177]
	{HST+COS spectra of the double white dwarf CSS~41177 place the secondary inside the pulsational instability strip}
\author[M.C.P. Bours et al.]
	{M.C.P. Bours$^{1}$\thanks{E-mail:m.c.p.bours@warwick.ac.uk},
	T.R. Marsh$^{1}$,
	B.T. G\"ansicke$^{1}$ and
	S.G.~Parsons$^2$. \\
	$^{1}$Department of Physics, University of Warwick, Coventry CV4 7AL, UK \\
	$^2$Departmento de F\'isica y Astronom\'ia, Universidad de Valpara\'iso, Avenida Gran Bretana 1111, Valpara\'iso 2360102, Chile }

\begin{document}

\date{Accepted .... Received ....; in original form ....}

\pagerange{\pageref{firstpage}--\pageref{lastpage}} \pubyear{2002}

\maketitle

\label{firstpage}

\begin{abstract}
We present Hubble Space Telescope + Cosmic Origins Spectrograph (HST+COS) data of the eclipsing double white dwarf binary CSS 41177. Due to the temperature difference between the two white dwarfs, the HST+COS far-ultraviolet data are dominated by the hot, primary white dwarf and allow us to precisely measure its temperature ($T_1$). Using eclipse observations, we also tightly constrain the temperature of the cooler secondary white dwarf ($T_2$). Our results, where $T_1$~=~22439~$\pm$~59~K and $T_2$~=~10876~$\pm$~32~K, with the uncertainties being purely statistical, place the secondary inside and close to the blue edge of the empirical instability strip for low temperature hydrogen-atmosphere white dwarfs. Dedicated high-speed photometry is encouraged to probe for the presence of pulsations, which will constrain the border of the instability strip as well as probe a new region of low gravity within the strip.
\end{abstract}


\section{Introduction}
CSS~41177, also known as SDSS~J100559.10+224932.3, was discovered to be an eclipsing binary containing a white dwarf during a search for transiting planets around white dwarf stars using Catalina Sky Survey (CSS) data \citep{Drake10}. Using an additional detection in the 2MASS survey, the authors put a weak constraint of spectral type M6 on the secondary star. However, given that the CSS light curve was not well sampled, they also mention that the data are consistent with a small faint object as the companion to the white dwarf. 

\citet{Parsons11} obtained higher quality photometry using the Liverpool Telescope \citep[LT,][]{Steele04} + RISE camera, which led to the detection of the secondary eclipse and the realisation that CSS~41177 is a double white dwarf binary with an orbital period of 2.78~hrs. CSS~41177 was the second eclipsing double white dwarf binary to be discovered, and it is still one of only five known binaries of this type \citep{Steinfadt10,Vennes11,Brown11,Kilic14}. In addition, CSS~41177 is the only one of these that is a double-lined spectroscopic binary, making it ideally suited for a detailed parameter study. \citet{Parsons11} combined the phase-resolved LT + RISE photometry with phase-resolved Gemini Multi-Object Spectrograph (GMOS) spectroscopy to determine orbital parameters and the white dwarf masses and radii.

In January 2012, higher signal-to-noise photometric data were obtained using the high-speed camera ULTRACAM \citep{Dhillon07}, which images in three bands simultaneously (Sloan Digital Sky Survey \citep[SDSS][]{York00} u$^{\prime}$g$^{\prime}$r$^{\prime}$ filters in this case), while it was mounted on the William Herschel Telescope and New Technology Telescope. Combined with Very Large Telescope (VLT) + X-Shooter spectroscopy with a time resolution of 5-6 minutes, the uncertainties on the masses and radii were substantially reduced \citep{Bours14a}. In addition, the changes in the relative eclipse depths in the three ULTRACAM bands constrained the primary white dwarf's temperature, while the differences in primary and secondary eclipse depths also constrained the secondary's temperature: $T_1$~=~24407(654)~K and $T_2$~=~11678(313)~K. Combined with the surface gravities from the ULTRACAM modelling, it was found that the secondary, cooler white dwarf was positioned very close to the empirical ZZ Ceti instability strip. Most white dwarfs found in this region of temperature - surface gravity parameter space show pulsations on timescales of 7-70 minutes. However, the ZZ Ceti strip is largely defined by a group of pulsating white dwarfs with surface gravities log~$g \sim$ 8~$\pm$~0.5 \citep{Gianninas06} and a handful of pulsators near log~$g$ = 6.5 \citep{Hermes12,Hermes13c}.  Because the cool, secondary white dwarf in CSS 41177 lies in between these two groups, near log~$g$ = 7.3, confirmation of its (non-)pulsating nature could help determine if the instability strip is continuous and if the current extrapolated empirical boundaries are correct, as predicted by theory \citep{VanGrootel13}. In our fits of the optical ULTRACAM light curves, the values of $T_1$ and $T_2$ were correlated, but at far-ultraviolet wavelengths the hotter primary white dwarf dominates the flux. Here we describe an improved parameter determination of CSS~41177 based on far-ultraviolet HST+COS observations.

\section{HST+COS observations} \label{sect:observations}
\begin{figure}
\includegraphics[]{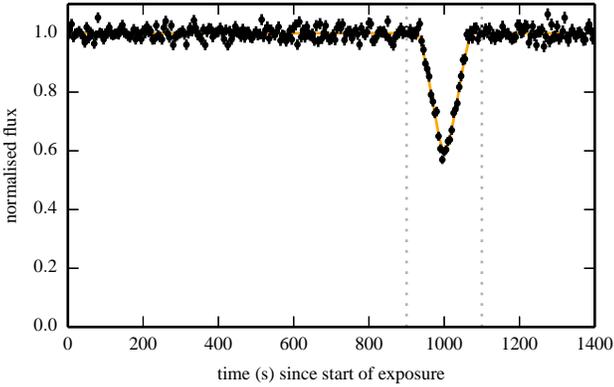}
\caption{HST+COS light curve of CSS~41177, displayed in 5~second bins, showing the eclipse of the hotter white dwarf (primary eclipse) and the model used to determine the mid-eclipse time (see Section~\ref{sect:observations}). The data between the dotted lines was excluded from the spectroscopic fits. }
\label{fig:hst_lc_eclipse}
\end{figure}

We obtained far-UV HST+COS \citep{Green12} data of CSS~41177 as part of Cycle 21, program ID 13421. CSS~41177 was observed during two consecutive HST orbits on May 28, 2014. Each orbit was split in two exposures to retain a good signal-to-noise ratio while minimising the effects of fixed pattern noise using all four FP+POS settings. The exposures obtained were 19, 20 and 2x24 minutes long and each used the G140L grating at a central wavelength of 1105~\AA. During the third exposure a primary eclipse occurred (cycle 7491, using the ephemeris from \citet{Bours14a}, with a mid-eclipse time $t_{\mathrm{MJD(UTC)}}$~=~56805.41665(34), equivalent to the barycentrically-corrected time $t_{\mathrm{BMJD(TDB)}}$~=~56805.41628(34), see Fig.~\ref{fig:hst_lc_eclipse}. % tmid[s] = 1001.1997775112737 +/- 29.127009

We removed the part of the exposure in which the eclipse occurred for the following analysis using \texttt{calcos} to filter out the data at 900 - 1100 s. We also removed geocoronal Lyman-$\alpha$ and O\textsc{i}~(1304~\AA) emission and interstellar absorption lines (see Section~\ref{sect:metallines}). The lower limit for our data is set at 1150~\AA, to exclude artificial features near the edge of the observable wavelength range. The upper limit is fixed at 1700~\AA, motivated by the decreasing sensitivity of COS and the therefore increasing difficulty of relative flux calibration at longer wavelengths.

\section{Results}
\subsection{Finding temperatures and surface gravities} \label{sect:method}
\begin{figure*}
\includegraphics[]{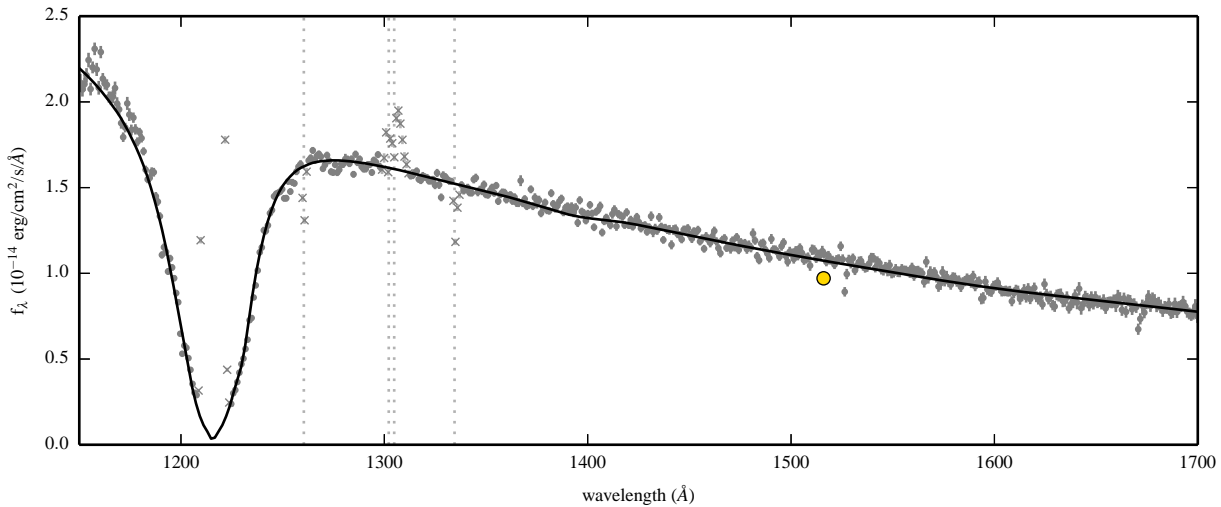}
\caption{HST+COS spectrum of CSS~41177 (grey dots, 1~\AA~bins; crosses denote data excluded from fits). The solid black line is the sum of the primary and secondary white dwarf model spectra corresponding to the best fit from the HST+ULTRACAM analysis described in Section~\ref{sect:method}. Also shown is the GALEX far-UV flux (yellow dot, errorbar too small to be seen) and the position of interstellar absorption lines (vertical grey dotted lines).}
\label{fig:hstfit_hstucam}
\end{figure*}

\begin{figure*}
\includegraphics[]{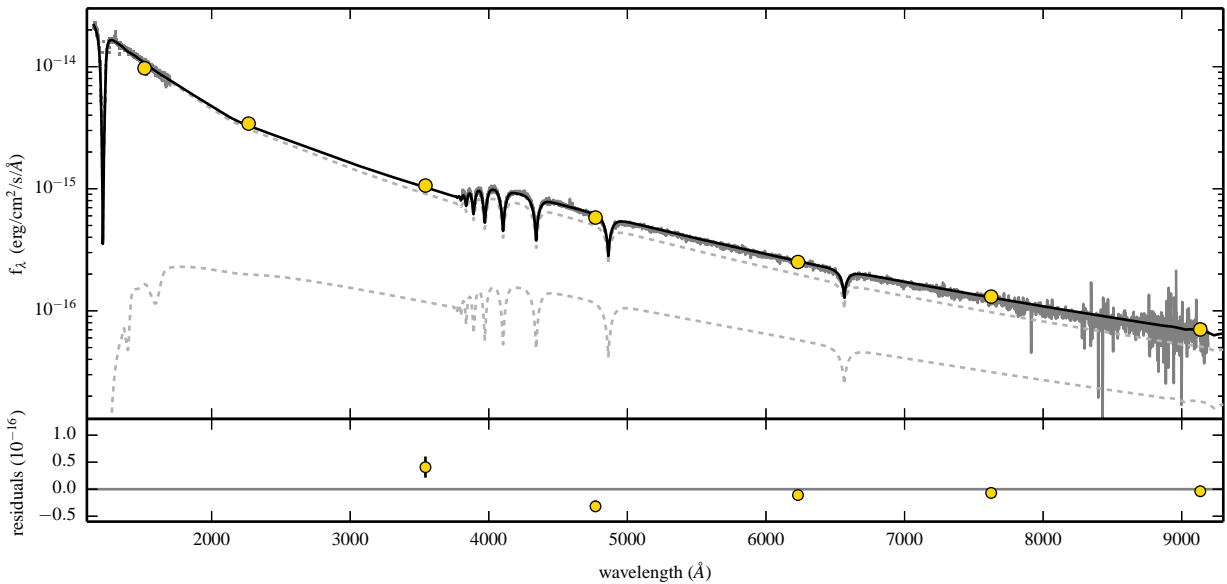}
\caption{\emph{Top panel:} model spectra for the double white dwarf binary CSS~41177 and individual white dwarfs (dotted grey lines). The HST+COS and SDSS spectra are shown in dark grey. Solid yellow dots indicate the GALEX far-UV, GALEX near-UV and SDSS ugriz fluxes (errorbars too small to be seen). \emph{Bottom panel:} residuals of the model CSS~41177 spectra folded through the SDSS filter curves with respect to the measured SDSS fluxes.}
\label{fig:spectrum_hstfit}
\end{figure*}

To determine the white dwarf temperatures, we fit the data with model spectra \citep{Koester10}, with a mixing length ML2/$\alpha$ = 0.8. However, simply fitting the HST+COS data with a single white dwarf model spectrum over predicts the flux from the hot white dwarf at optical wavelengths. Because the secondary white dwarf starts contributing to the total flux at $\lambda >$ 1400~\AA, the slope of the HST+COS spectrum is shallower than would be the case if only the primary white dwarf is visible in this regime. As a result, the temperature we would find for the primary white dwarf is too low. Therefore we use an iterative process which, after the first step, includes the secondary white dwarf's contribution to the far-UV flux in order to find a more accurate temperature for the primary white dwarf.

In the first step we fit the HST+COS spectra with a single white dwarf model spectrum using a Markov-chain Monte Carlo (MCMC) analysis using the python package \texttt{emcee} \citep{ForemanMackey13}. This MCMC fit includes as free parameters the primary white dwarf's temperature ($T_1$) and surface gravity (log~$g_1$), the reddening towards CSS~41177 ($E(B-V)$), and a scale factor to allow for the unknown distance to the binary. The surface gravity is constrained by a Gaussian prior at log~$g_1$ = 7.32 $\pm$ 0.02 \citep{Bours14a} and the reddening by a Gaussian prior at $E(B-V)$ = 0.0292 $\pm$ 0.0009 \citep{Schlafly11}.

We then fitted the ULTRACAM light curve data as presented and discussed in \citet{Bours14a}, but now with a prior on the temperature of the primary white dwarf $T_1$ based on the results from the MCMC analysis of the HST+COS data described above. Briefly, the ULTRACAM data include twelve primary and nine secondary eclipses, all observed simultaneously in the SDSS u$^{\prime}$g$^{\prime}$r$^{\prime}$ filters. The eclipse depths in these three bands constrain the relative temperatures of the white dwarfs. The free parameters in this fit are $T_1$ and $T_2$, the radii relative to the binary's semi-major axis $R_1/a$ and $R_2/a$, the radial velocity amplitudes $K_1$ and $K_2$ (constrained by X-Shooter spectroscopy also presented in \citealt{Bours14a}), the zero-point of the ephemeris $T_0$, the orbital period and inclination $P_{\mathrm{orb}}$ and $i$, the reddening $E(B-V)$ and an offset $\delta$ from phase 0.5 of the secondary eclipse that allows for a R\o mer delay or eccentricity of the orbit. The tight prior on $T_1$ given by the HST+COS data leads to a much more precise temperature for the secondary white dwarf $T_2$ as well.

This, in turn, allows us to refit the HST+COS data while accounting for `contamination' to the UV-flux by the secondary white dwarf by subtracting a reddened model spectrum from the data, with values $T_2$ and log~$g_2$ fixed at the mean values from the ULTRACAM light curve fit. The new value found for $T_1$ then becomes a more accurate prior in the light curve fits. By including the contribution of the secondary white dwarf, the primary white dwarf's temperature increases by $\sim$~150 K after the first iteration, and $\sim$~10 K after the second iteration. The latter is well within the statistical uncertainty and eliminates the need for further iterations.

The converged results are listed in Table~\ref{tab:parameters}. We have experimented with placing the long wavelength cut on the HST+COS spectra at 1500~\AA, which decreases the primary white dwarf's temperature by $\sim$~150 K, roughly illustrating the magnitude of the systematic uncertainties. The uncertainties in Table~\ref{tab:parameters} should be considered as statistical uncertainties only.

The HST+COS data and the best synthetic model for the double white dwarf from the MCMC analysis are shown in Fig.~\ref{fig:hstfit_hstucam}. Fig.~\ref{fig:spectrum_hstfit} shows the model spectra for the individual white dwarfs as well, and covers the far-UV through to the near-IR wavelength range. The CSS~41177 models are the sum of two white dwarf model spectra, accounting for the difference in their surface area. The parameters (T$_{\mathrm{eff}}$, log~$g$) for the primary white dwarf's model spectrum are those from the best fit in the last iteration to the HST+COS data, and the parameters for the secondary white dwarf are taken from the corresponding model to the ULTRACAM light curves.

The CSS 41177 model spectrum is in good agreement with the HST+COS data as well as with the SDSS data at all wavelengths. The slight over- and under-predictions of the (AB corrected) flux that can be seen in the residuals in the bottom panel of Fig.~\ref{fig:spectrum_hstfit} amounts to 2-3 times the uncertainty on the measured flux or, equivalently, a few \%. Because the SDSS uncertainties do not include systematic uncertainties, we do not think this difference is significant.

\begin{table}
\begin{center}
\caption{White dwarf parameter results from the MCMC analyses performed on the HST+COS and ULTRACAM data. For comparison, the results from \citet{Bours14a} based on the ULTRACAM light curves alone are shown as well ($E(B-V)$ was fixed to the mean value from \citealt{Schlegel98}). Numbers in parentheses indicate uncertainties in the last digit(s). }
\label{tab:parameters}
\begin{tabular}{p{2.5cm}p{2.5cm}p{2.2cm}}
\hline
parameter                  & HST+ULTRACAM   &  ULTRACAM             \\
\hline                     
$T_1$ (K)                  & 22439(59)      &  24407(654)           \\
$T_2$ (K)                  & 10876(32)      &  11678(313)           \\
$E(B-V)$ (mag)             & 0.0292(9)      &  0.0339 (fixed)       \\
log$(g_1)$                 & 7.322(15)      &  7.321(15)            \\
log$(g_2)$                 & 7.305(11)      &  7.307(11)            \\
$R_1/a$                    & 0.02508(22)    &  0.02510(22)          \\
$R_2/a$                    & 0.02340(28)    &  0.02332(28)          \\
minimum $\chi^2$           & 44465          &  44457                \\
degrees of freedom         & 37745          &  43351                \\
\hline
\end{tabular}
\end{center}
\end{table}

\section{Metal lines in the HST+COS spectra} \label{sect:metallines}
The HST+COS spectra reveal a handful of absorption lines. The position of these lines coincides with known interstellar absorption features: Si\textsc{ii} 1260~\AA; O\textsc{i} 1302,1304~\AA; C\textsc{ii} 1334/5~\AA. Given that the lines also do not shift position between the four individual exposures we conclude that they are indeed of interstellar origin. The Si\textsc{ii} 1265~\AA~excited state is not detected, further corroborating the interstellar nature of the metal lines.

\section{A pulsating secondary white dwarf?}
\begin{figure}
\includegraphics[width=0.49\textwidth]{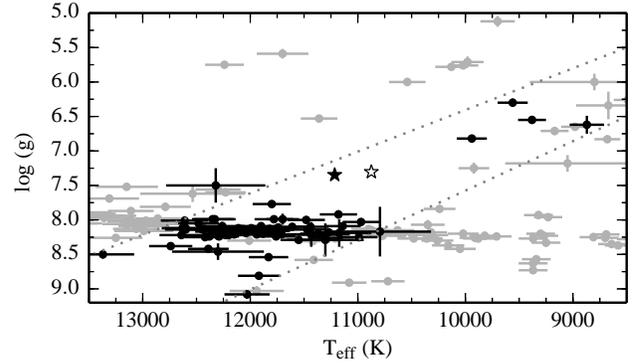}
\caption{ZZ Ceti diagram, with the filled, black star showing the position of the secondary white dwarf in CSS~41177 at the overestimated values that would have been recovered if the parameters were found by fitting the secondary white dwarf's spectra with 1D model atmospheres \citep{Tremblay13b}. The open star indicates the direct values found from our fit -- see text. The black and grey dots indicate confirmed pulsating and non-pulsating white dwarfs respectively, and the dotted lines are empirical boundaries for the instability strip \citep{Gianninas14}. }
\label{fig:logg_vs_teff_hstfit}
\end{figure}

Depending on their atmospheric composition, white dwarfs experience non-radial gravity-mode pulsations at different epochs in their cooling curves. Based on extrapolating observations of white dwarfs with lower and higher surface gravities, this instability should occur close to T$_{\mathrm{WD}}$ = 11000 K for white dwarfs with a mass of M$_{\mathrm{WD}}$ $\sim$ 0.3 M$_{\odot}$ and hydrogen-rich atmospheres (DA) \citep[see e.g.][]{Bergeron95,Gianninas14}. In the log~$g$ - T$_{\mathrm{eff}}$ plane this region is called the ZZ Ceti instability strip.

The values of the secondary white dwarf's temperature and surface gravity from both our analyses are shown in Fig.~\ref{fig:logg_vs_teff_hstfit}, where known pulsators \citep{Gianninas11, Hermes12, Hermes13b, Hermes13c, Hermes14a, Pyrzas15} and non-pulsators at 10~mmag \citep{Gianninas11, Steinfadt12, Hermes12, Hermes13c, Hermes13a} are also indicated. The values used for low-mass white dwarfs are the updated atmospheric parameters from \citet{Gianninas14}, who used the latest model spectra, with a 1-dimensional mixing length theory to approximate convection, to fit the white dwarf Balmer lines. It has been known for some time that fitting Balmer lines with 1D models over predicts surface gravities and temperatures in the regime where white dwarf atmospheres become convective \citep[T$_{\mathrm{eff}} <$ 13000 K;][]{Tremblay10, tremblay11b}.

The method presented in this paper does not rely on fitting Balmer lines, but rather uses the continuum of the spectral energy distribution and the ULTRACAM light curves to find the temperature and surface gravity of the secondary white dwarf in CSS~41177. These results are likely close to the true atmospheric parameters, and therefore presumably similar to results that would be obtained if it was possible to fit the secondary's Balmer lines directly with the latest 3D model spectra. For this reason, we correct our results in Fig.~\ref{fig:logg_vs_teff_hstfit} using the 3D to 1D correction given by \citet{Tremblay13b}, to facilitate direct comparison with other data points in this figure. The filled star indicates the position in the ZZ Ceti strip that would have been recovered if fits to the secondary star's Balmer lines were performed with 1D model atmospheres. The open star shows the atmospheric parameters found with our method, also listed in Table~\ref{tab:parameters}. Note that the average of the 3D models is performed over constant Rosseland optical depth, which produces very similar results to full 3D spectral synthesis at all wavelengths for these temperatures and surface gravities \citep{tremblay11b}.

Our results place the secondary white dwarf inside the empirical instability strip, just under 350~K from the blue edge. With the available ULTRACAM data, pulsations with a relative amplitude exceeding 0.5\% in the light of the secondary (i.e. correcting for the 80\% contribution of the hot primary) were ruled out \citep{Bours14a}. However, as noted in that paper, the ULTRACAM data were not ideally suited for finding pulsations. The observations targeted the eclipses, and mostly covered only brief stretches (5-10 minutes) of out-of-eclipse data. Typically, pulsation periods are between $\sim$~7-70 minutes \citep{Hermes12,Hermes13c}, and increase with decreasing surface gravity and temperature. Therefore the ULTRACAM observations are too short. Furthermore, pulsating white dwarfs can have amplitudes $<$ 0.5\% \citep{Mukadam06}. This might explain why we have not been able to detect any pulsations. New, dedicated observations will reveal whether pulsations are indeed present or not.

Note that we have assumed a prior for the reddening at $E(B-V)$ = 0.0292 $\pm$ 0.0009 in our MCMC analyses, which is the total estimated reddening in the direction of CSS~41177. Since the reddening affects the flux at shorter wavelengths more than the flux at longer wavelengths, a smaller $E(B-V)$ results in a shallower slope of the HST+COS spectra. This translates to a lower temperature for the primary white dwarf, and, through the ULTRACAM light curves, also in a lower temperature of the secondary white dwarf. Therefore any decrease in the reddening will push the secondary white dwarf further into the instability strip.

\section{Conclusions}
We have presented HST+COS spectra of the eclipsing double white dwarf binary CSS~41177. By fitting the far-UV data with model spectra, these allow us to precisely measure the primary white dwarf's temperature: $T_1$~=~22439 $\pm$ 59 $\pm$ 150 K. Combined with the eclipse data in previously presented u$^{\prime}$g$^{\prime}$r$^{\prime}$ ULTRACAM data this puts a very tight constraint on the secondary temperature as well: $T_2$~=~10876 $\pm$ 32 $\pm$ 150 K.

The ULTRACAM light curves, combined with constraints from X-Shooter time-series spectroscopy also presented in \citet{Bours14a}, also constrain the surface gravity: log~$g_2$~=~7.305 $\pm$ 0.011. These results place the secondary white dwarf inside the pulsational instability strip for white dwarfs with hydrogen-rich atmospheres, and in particular in the gap between the canonical log~$g$ $\simeq$ 8 white dwarfs and the extremely low mass white dwarfs at log~$g$ $\simeq$ 6.5. The previously presented ULTRACAM data include only short stretches of out-of-eclipse data, in which we were unable to detect pulsations. Given the results presented here and the predictions made by theory \citep{VanGrootel13}, dedicated high-speed photometric observations resulting in higher signal-to-noise ratio data may still reveal the presence of pulsations. This white dwarf is particularly interesting because it is positioned in an as-yet-unexplored part of the ZZ Ceti diagram, as well as being close to the edge of the instability strip. Determining if it pulsates or not will help determine where the border of the instability strip is and whether the strip is continuous from high mass to extremely low mass white dwarfs. \\

\footnotesize{
\noindent\textbf{Acknowledgements} 
We kindly thank the referee for the swift review of this manuscript. Based on observations made with the NASA/ESA Hubble Space Telescope, obtained at the Space Telescope Science Institute, which is operated by the Association of Universities for Research in Astronomy, Inc., under NASA contract NAS 5-26555. These observations are associated with program \#13421. TRM acknowledges financial support from STFC under grant number ST/L000733/1. The research leading to these results has received funding from the European Research Council under the European Union's Seventh Framework Programme (FP/2007-2013) / ERC Grant Agreement n. 320964 (WDTracer). SGP acknowledges financial support from FONDECYT in the form of grant number 3140585. }

\bibliographystyle{mn_new}

\bsp

\label{lastpage}

\end{document}